
\input harvmac
\noblackbox
\def\inbar{\,\vrule height1.5ex width.4pt depth0pt}
\font\cmss=cmss10
\def\BZ{\relax\hbox{\cmss Z\kern-.4em Z}}
\def\IC{\relax\hbox{$\inbar\kern-.3em{\rm C}$}}
\def\IP{\relax{\rm I\kern-.18em P}}
\def\CP#1{{\IP}^{#1}}

\lref\swit{N. Seiberg and E. Witten, ``Electromagnetic Duality,
Monopole Condensation and Confinement in $N{=}2$ Supersymmetric
Yang-Mills Theory'',
Nucl. Phys. {\bf B426} (1994) 19--52.}
\lref\west{P.C. West, Introduction to Supersymmetry and Supergravity,
 World Scientific, Singapore, 1986.}
\lref\schoen{C. Schoen, ``On Fiber Products of Rational Elliptic
Surfaces with Section'', Math. Zeit. {\bf 197} (1988) 177--199.}
\lref\werner{J. Werner and B. van Geemen, ``New Examples of
Threefolds with $c_1=0$'', Math. Zeit. {\bf 203} (1990) 211--225.}
\lref\clemens{C.H. Clemens,
``Double Solids'',  Adv. Math. {\bf 47} (1983) 107--230.}
\lref\friedman{R. Friedman,
``Simultaneous Resolution of Threefold Double Points'',
Math. Ann. {\bf 274} (1986) 671--689.}
\lref\hirzebruch{F. Hirzebruch (notes by J. Werner),
``Some Examples of Threefolds
with Trivial Canonical Bundle'', in Gesammelte Abhandlungen, Band II,
Springer-Verlag, Berlin-New York, 1987, pp.~757--770.}
\lref\tianyau{G. Tian and S.-T. Yau,
``Three Dimensional Algebraic Manifolds with $c_1=0$ and $\chi=-6$'',
in Mathematical Aspects of String Theory (S.-T. Yau, ed.),
World Scientific, Singapore, 1987, pp.~629--646.}
\lref\morrison{D.R. Morrison, ``Through the Looking Glass'',
Lecture at CIRM conference, Trento (June, 1994), to appear.}
\lref\rgpmb{B.R. Greene and M.R. Plesser, ``An Introduction to
Mirror Manifolds'',
in Essays on Mirror Manifolds (S.-T. Yau, ed.), International Press,
Hong Kong, 1992, pp.~1--30.}
\lref\agmmult{P.S. Aspinwall, B.R. Greene and D.R. Morrison,
``Multiple Mirror Manifolds and Topology Change in String Theory'',
Phys. Lett. {\bf B303} (1993) 249--259.}
\lref\wittenphases{E. Witten, ``Phases of $N{=}2$ Theories in Two Dimensions'',
Nucl. Phys. {\bf B403} (1993) 159--222.}
\lref\gp{B.R. Greene and M.R. Plesser, ``Duality in Calabi--Yau Moduli
Space'',
Nucl. Phys. {\bf B338} (1990) 15--37.}
\lref\ferr{A. Ceresole, R. D'Auria, S. Ferrara and A. Van Proeyen,
``Duality Transformations in Supersymmetric Yang-Mills Theory Coupled
to Supergravity, '' hep-th/9502072.}
\lref\bhole{G. Horowitz and A. Strominger, ``Black Strings and
$p$-branes,'' Nucl. Phys. {\bf B360} (1991) 197--209.}
\lref\lefschetz{S. Lefschetz, L'Analysis Situs et la G\'eom\'etrie
Alg\'ebrique, Gauthier-Villars, Paris, 1924; reprinted in
Selected Papers, Chelsea, New York, 1971, pp.~283--439.}
\lref\agm{P.S. Aspinwall, B.R. Greene and D.R. Morrison, ``Calabi--Yau
Moduli Space, Mirror Manifolds and Spacetime Topology Change
in String Theory'', Nucl. Phys. {\bf B416} (1994) 414--480.}
\lref\reid{M. Reid, ``The Moduli Space of $3$-folds with $K{=}0$ May
Nevertheless Be Irreducible'', Math. Ann. {\bf 278} (1987) 329--334.}
\lref\texasi{ P. Candelas, A.M. Dale, C.A. L\"utken, and R. Schimmrigk,
``Complete Intersection Calabi--Yau Manifolds'', Nucl. Phys.
{\bf B298} (1988) 493--525.}
\lref\texasii{P.S. Green and T.  H\"ubsch,
``Possible Phase Transitions Among Calabi--Yau Compactifications'',
Phys. Rev. Lett.  {\bf 61} (1988) 1163--1166;
``Connecting Moduli Spaces of Calabi--Yau Threefolds'', Comm. Math.
Phys.  {\bf 119} (1988) 431--441.}
\lref\texasiii{P. Candelas, P.S. Green, and T. H\"ubsch,
``Finite Distance Between Distinct Calabi--Yau Manifolds'',
Phys. Rev. Lett.  {\bf 62} (1989) 1956--1959;
``Rolling Among Calabi--Yau Vacua'', Nucl. Phys.
{\bf B330} (1990) 49--102.}
\lref\cdcon{ P. Candelas and X.C. de la Ossa,
``Comments on Conifolds'', Nucl. Phys. {\bf B342} (1990) 246--268.}
\lref\mbh{A. Strominger, ``Massless Black Holes and Conifolds in String
Theory'', hep-th/9504090.}
\lref\bbgms{K. Becker, M. Becker, B.R. Greene, D.R. Morrison and A. Strominger,
to appear.}
\lref\dlwp{B. de Wit, P. Lauwers and A Van Proeyen, ``Lagrangians of $N{=}2$
Supergravity-Matter Systems'', Nucl. Phys. {\bf B255} (1985) 569--608.}
\lref\ascmp{A. Strominger, ``Special Geometry'', Comm. Math. Phys. {\bf 133}
(1990) 163--180.}
\lref\nati{N. Seiberg, ``Observations on the Moduli Space of Superconformal
Field Theories'', Nucl. Phys. {\bf B303}, (1988) 286--304.}
\lref\serg{S. Cecotti, S. Ferrara and L. Girardello,
``Geometry of Type II Superstrings and the Moduli of
Superconformal Field Theories'', Int. J. Mod. Phys. {\bf 4} (1989) 2475--2529.}
\lref\udual{P.S. Aspinwall and D.R. Morrison, ``U-Duality and Integral
Structures'', CLNS-95/1334, to appear.}

\ifx\answ\bigans
\Title{\vbox{\baselineskip12pt
\hbox{CLNS-95/1335}\hbox{hep-th/9504145}}}
{\vbox{\centerline{\bf{ BLACK HOLE CONDENSATION AND }}
\vskip2pt\centerline{\bf{THE UNIFICATION OF STRING VACUA}}}}
\else
\Title{\vbox{\baselineskip12pt
\hbox{CLNS-95/1335}\hbox{hep-th/9504145}}}
{\vbox{\centerline{\bf{ BLACK HOLE CONDENSATION AND
THE UNIFICATION OF STRING VACUA}}}}
\fi

{
\baselineskip=12pt
\centerline{ Brian R. Greene }
\smallskip
\centerline{\sl F.R. Newman Laboratory of Nuclear Studies}
\centerline{\sl Cornell University}
\centerline{\sl  Ithaca, NY  14853}
\bigskip
\centerline{ David R. Morrison\footnote{$^\dagger$}{On leave from:
Department of Mathematics, Duke University, Durham, NC 27708-0320} }
\smallskip
\centerline{\sl Department of Mathematics}
\centerline{\sl Cornell University}
\centerline{\sl  Ithaca, NY  14853}
\bigskip
\centerline{\it and}
\smallskip
\centerline{ Andrew Strominger }
\smallskip
\centerline{\sl Department of Physics}
\centerline{\sl University of California}
\centerline{\sl Santa Barbara, CA 93206-9530}

\bigskip
\centerline{\bf Abstract}
It is argued that black hole condensation can occur at conifold
singularities in the moduli space of type II Calabi--Yau string vacua. The
condensate signals a smooth transition to a  new Calabi--Yau space with
different Euler characteristic and Hodge numbers. In this manner string theory
unifies the  moduli spaces of many or possibly all
Calabi--Yau vacua. Elementary string states and black holes
are smoothly interchanged under the
transitions, and therefore cannot be invariantly distinguished.
Furthermore,
the transitions establish the existence of mirror symmetry for many
or possibly all Calabi--Yau
manifolds.
}

\Date{4/95}

\newsec{Introduction and summary}

In ten dimensions, there is a small number of consistent string theories. When
first discovered, this near-uniqueness
raised the hope that, despite the inherent difficulties in
extrapolating
from the
Planck scale to the weak scale, it might be possible to obtain testable
predictions
from
string theory. This hope was greatly diminished by the discovery of a plethora
of four-dimensional string vacua.
In order to compute the low-lying spectrum and
couplings
it is
apparently first necessary to choose among many thousands of Calabi--Yau spaces
(or more
general conformal field theories).  The process of compactification appeared to
ruin the uniqueness of ten-dimensional string theory, along with the predictive
power it entailed.

In this paper we will argue that the situation is in fact much better than it
appears.
In the context of type II strings, we will see that many, and possibly all, of
these
Calabi--Yau vacua are in fact different branches of a vastly larger
``universal''
moduli
space.\foot{
A previous step in this direction was taken in \refs{\agm,\wittenphases}, where
string compactifications which
are topologically distinct as Calabi--Yau spaces but not as conformal field
theories
were smoothly connected.}
The branches are connected in a smooth and calculable manner by black hole
condensation
which can occur at ``conifold'' points of the moduli space. This condensation
can not be
described in the language of conformal field theory, and so is not constrained
to
preserve
quantities  such as the number of light generations which are topological
invariants of
conformal field theory. Thus the number of distinct four-dimensional
string vacua is much
smaller than previously suspected. Indeed, it is conceivable that there is
a unique four-dimensional string vacuum.

Of course even if the goal of tying together all type II string vacua is
realized, we
are still quite far from making testable predictions. One must extend these
ideas to
theories with $N=1$ supersymmetry in four dimensions, and then understand how a
superpotential is generated which lifts the continuous vacuum degeneracy  and
breaks $N=1$
down to $N=0$. Even then there is no guarantee that there is a unique or
small number of vacua.
Nevertheless we feel the time may be ripe for progress on all
these problems.

It has long been known in the mathematics literature
\refs{\clemens\friedman\hirzebruch{--}\tianyau} that it is possible
to
travel from one Calabi--Yau to another by degenerating certain three-cycles and
then
blowing them back up as two-cycles, thereby changing the Hodge numbers. This
process enables one to glue together Calabi--Yau moduli spaces along the
subspaces corresponding to conifolds.
Indeed it has been conjectured \reid\ that all Calabi--Yau
spaces are connected
in this fashion. The relevance of these results to string theory was advocated
in a
series of prescient papers \refs{\texasi\texasii\texasiii{--}\cdcon}.
However the construction, at the level of both mathematics
and conformal field theory, is  singular, and no
proposal was made for how string theory might physically
implement the transition from one
Calabi--Yau to its
neighbor. The purpose of the present work is to describe such a mechanism.

The key to understanding this transition is in a recent resolution \mbh\ of
simple conifold singularities in
type II string theory. Near a conifold, the moduli space metric becomes
singular. At the
same time,
string theory contains black hole hypermultiplets which are degenerating to
zero mass.
It was shown \mbh\ that the Wilsonian effective field theory including the
light
black holes
is smooth near the conifold, and that integrating out the light black holes
reproduces the
singularity.

The singularities of the conifolds which glue together Calabi--Yau moduli
spaces are more
complicated than
the simple type analyzed in \mbh. In this paper we shall again find that these
singularities are
resolved by light
black hole hypermultiplets, but there are in general many such hypermultiplets.
The
potential $V$ for these hypermultiplets is determined by $N=2$ supersymmetry.
$V$ has flat directions, along which the black holes can condense and give
masses to
vector multiplets. In this way one discovers a new branch of the moduli space
with
different
numbers of hypermultiplets and vector multiplets. In a IIB string theory the
number of massless
vector multiplets (hypermultiplets) is $h_{21}$ ($h_{11}+1$) and this new
branch
corresponds to a topologically distinct Calabi--Yau space. In this paper we
analyze
only the
transition from the quintic in $\CP4$ with  Hodge numbers
$(h_{21},h_{11})=(101,1)$, to a variety in $\CP4\times \CP1$,with Hodge numbers
$(86,2)$, but our construction clearly generalizes.
Further analysis will appear in a forthcoming publication \bbgms.

Our construction has  a number of implications that are worth emphasizing at
the outset. First, as mentioned, type II string vacua which were previously
thought
to be disjoint are now seen to fit together into a connected web with string
physics smoothly interpolating from one component to another. Second, as
we move along such an interpolating path, the spacetime background of
our string theory undergoes a drastic
change in topology. Unlike the spacetime topology
change of \refs{\agm,\wittenphases}, in which Hodge numbers stay fixed
while more
subtle topological invariants (such as the intersection form) change, here
we find that string theory is perfectly smooth even as the Hodge numbers jump.
Third, using these results we can vastly extend the mirror symmetry
construction
of \gp. Namely, on general grounds, once one proves the existence of mirror
symmetry for one point in a given moduli space, via deformation arguments one
can conclude the existence of mirror symmetry at all other points lying in
the same connected component. This was used in \gp, for instance, to argue for
mirror symmetry throughout a Calabi--Yau moduli space so long as it contains
a minimal model point at which an explicit mirror partner
exists. Now we see that our
deformation
arguments are not limited to a single Calabi--Yau moduli space, but rather
extend
to all Calabi--Yau manifolds connected by conifold transitions; the latter
includes essentially
all known Calabi--Yau manifolds.\foot{In \refs{\texasii,\texasiii} it
was established that all simply-connected
Calabi--Yau manifolds {\it known at that
time}\/ could be connected in this way.  (It is easiest to deal with multiply
connected
Calabi--Yau's by working with their covering spaces.)
Since the time of \refs{\texasii,\texasiii} a number of
new constructions of Calabi--Yau manifolds have been proposed, and
while it is clear that most of these can be connected, no systematic
study has been made of this question.}
Finally,
our construction  provides food for thought on the fascinating interplay
between strings
and black holes. A degenerating black hole hypermultiplet is reinterpreted as a
fundamental
string excitation (corresponding to a modulus) after crossing the transition.
Thus there can
be no fundamental distinction between strings and black holes: they smoothly
transform
into one another.

\newsec{Classical structure of conifolds}

For concreteness, we focus on a particular example in this paper,
although our analysis is general.
The moduli space $\cal M$ of all
complex structures on a quintic threefold $X$ can be described in
terms of the defining equations of the quintics: the general
such equation takes the form of a homogeneous quintic polynomial
\eqn\quintic{c_0\, x_0^5 + \cdots + c_{125}\, x_4^5 = 0}
and involves 126 coefficients.  Some 25 of these are redundant
due to the action of
$\mathop{\hbox{\rm GL}}(5)$ on the
homogeneous coordinates $[x_0,\dots,x_4]$, leaving 101 independent parameters
in the moduli space.

For general values of $(c_0,\dots,c_{125})$, the equation
\quintic\ defines a nonsingular Calabi--Yau manifold $X_c$ in $\CP4$, but
at certain special values (along a set $\Delta$ of complex codimension
$1$ in the parameter space), the solution set of \quintic\ becomes singular.
These singularities were analyzed many years ago by Lefschetz \lefschetz,
who showed that:

1) the ``generic'' singular space $X_c$, $c\in\Delta$ has a single
node,\foot{We distinguish between the singular points of the
Calabi--Yau---here called {\it nodes}---and the points in the moduli
space which label such Calabi--Yau's---called {\it conifold points}.}
i.e., a singular point with local equation $\sum_{i=1}^4 y_i^2=0$,

2) the singular point determines a ``vanishing cycle'' $\gamma\in H_3(X,Z)$
which shrinks to zero size when the singularity is approached,\foot{More
precisely, the period integrals over $\gamma$ vanish in the limit.} and

3) the homology of the Calabi--Yau manifolds
undergoes a monodromy transformation
\eqn\monodromy{\delta \mapsto \delta + \langle\delta|\gamma\rangle\,\gamma}
upon transport around a loop in $\cal M$ encircling the singular locus
$\Delta$.  (Here,
$\langle \delta |\gamma \rangle$ denotes the number of (oriented)
intersections of
$\delta$ with $\gamma$.)  If all of the singular points on $X_c$ are nodes,
then $X_c$ is  called
a ``Calabi--Yau conifold'' (so named \texasii\ because of the conical
nature of the singularities).

If we introduce a holomorphic $3$-form $\Omega$ (depending on the moduli),
and use the ``periods'' $\int_\delta\Omega$ of $\Omega$ to describe the
complex structure, we find that some of the periods become multiple-valued
near the singular locus.  In fact, if we let
\eqn\mainperiod{Z\equiv \int_\gamma\Omega}
be the period corresponding to the vanishing cycle, then we find
that the singular locus $\Delta$ in $\cal M$
is locally described by $Z=0$, and
that other periods must take the form
\eqn\otherperiods{\int_\delta\Omega \sim {1 \over 2 \pi i} \langle\delta
|\gamma\rangle\,Z\ln Z
+ \hbox{\rm (single-valued function)}}
near $Z=0$ in order to have  the correct monodromy property.

Let us now consider the set of quintic conifolds in $\CP4$ which have $k$
singular points.  Since asking for a single node places a single condition
on
the parameters,  one's initial expectation is that asking for a conifold
with $k$ singular
points will place $k$ conditions on the parameters, leading
to a locus of complex codimension $k$ in $\cal M$.  When this is true,  the
generic
point of that locus will locally be an intersection of $k$ hypersurfaces,
meeting transversally.  Near the intersection of all of these, there
will be $k$ different monodromy transformations with vanishing cycles
$\gamma^1$, \dots, $\gamma^k$, and the periods
must take the form
\eqn\asympt{\int_\delta\Omega \sim {1 \over 2 \pi i}
\sum_{a=1}^k\langle\delta|\gamma^a\rangle\,Z^a\ln Z^a
+ \hbox{\rm (single-valued function)}}
near $Z^1=\cdots=Z^k=0$, where $Z^a\equiv \int_{\gamma^a}\Omega$ are among
the local coordinates near the intersection.

However, in the example which we will study in detail there are
$k=16$ singular points on the conifold which impose only $15$ conditions on
the parameters.  In fact, the vanishing cycles in our example in the next
section
will satisfy the homology relation
\eqn\relation{\sum_{a=1}^{16}\gamma^a=0 ,}
which implies a corresponding relation among the periods:
\eqn\perrel{\sum_{a=1}^{16} Z^a=0 .}
The locus of conifolds with (at least) $15$ singular points is described by
$Z^1=\cdots=Z^{15}=0$, but since the hypersurface $Z^{16}=0$
passes through this locus (as a consequence of \perrel), the
$16^{\hbox{\ninerm th}}$ point is also present, without imposing
any further conditions.  This is clearly a special property of
the particular collection of $16$ points which we are considering.

In this situation, we have $16$ monodromy transformations near a locus
of codimension $15$.  The asymptotic form \asympt\ still holds
near $\{Z^a=0\}$, but
the $\gamma^a$ and $Z^a$ are related by \relation\ and \perrel.
In particular, only a subset of the $Z^a$'s can be included among
a list of local coordinates near the intersection locus.

\newsec{The example}

The example we study first appeared in the physics literature in
\texasiii. We will describe this example precisely in what follows, but
first let us emphasize the basic idea.  We start with a smooth quintic
in $\CP4$  and follow a path in its complex structure moduli space leading
us to a conifold with 16 singular points. Each of these singular points can
be described locally as a cone over an $S^2 \times S^3$ base. We resolve such
singularities by cutting out
a neighborhood of the singular point and gluing in
 a smooth space (of real dimension 6) whose boundary
agrees with that of the extracted set,
namely $S^2 \times S^3$.
 For the singular points of a conifold this can
be done in two ways: $i$) glue in $B^3 \times S^3$ or $ii$) glue in $S^2 \times
B^4$.
The former is a deformation of the conifold back into the
quintic moduli space, by giving positive volume
to the shrunken three-cycles. The latter is a small resolution of the conifold
by
giving positive area to the  $S^2$'s. Effectively, the small resolution
replaces
previous $S^3$'s with $S^2$'s and thereby changes the Hodge numbers, and hence
topology, of the Calabi--Yau space. (In fact, there are generally two distinct
ways of performing the small resolution that differ by a flop. This played
a key role in \refs{\agm,\wittenphases} but is not of central importance here.)
Our goal is to understand how type II string theory behaves as we attempt
to pass from the smooth quintic, through the conifold, and on to its
topologically distinct small resolution. To do so, we first recast the present
discussion into a more concrete form.

Consider  the set of quintics in $\CP4$ which contain a fixed
$\CP2$, say the one with $x_3=x_4=0$. We do this because,
as we shall see, the 16 singular points referred to above can be made
to all reside on this $\CP2$. The defining equation of
such a Calabi--Yau space must not contain any of the 21 monomials
$x_0^5$, \dots, $x_2^5$ which involve only $x_0$, $x_1$ and $x_2$;
there are therefore 105 parameters in the defining equation.
On the other hand, the number of redundancies has decreased,
since we are only free to use the subgroup of $\mathop{\hbox{\rm GL}}(5)$
which fixes the $\CP2$, i.e., matrices $(a_{jk})$ for which the
coefficients $a_{jk}$ vanish when $j=3, 4$ and $k=0, 1, 2$.
That subgroup has dimension 19, so the
total number of effective parameters is 86, a set of codimension $15$
in the original 101-dimensional space $\cal M$.

If we write the defining polynomial of such a quintic in the form
\eqn\split{f(x)=x_3\, g(x)+x_4\, h(x)}
where $g(x)$ and $h(x)$ are polynomials of degree $4$,
then by considering the partial derivatives of $f$ with respect to
$x_3$ and $x_4$ it becomes apparent that the quintic must
be singular along the set $\{x_3=x_4=g(x)=h(x)=0\}$ which consists
of the sixteen points of intersection of $g$ and $h$ within $\CP2$.
When $g$ and $h$ are generic, these singularities are simply
nodes.  That is, we have defined a Calabi--Yau {\it conifold}\/
rather than a Calabi--Yau {\it manifold}.

The vanishing cycles of these singular points are easy to locate.
To do so, consider a neighborhood of a given singular point on the
conifold that we obtain by
intersecting the quintic with a ball in $\CP4$, i.e., a $B^8$. As discussed in
\refs{\texasiii,\cdcon},
this intersection is a cone with base being $S^2 \times S^3$.
Therefore,
if we remove $16$ such small balls
from $\CP4$ about the $16$ singular points, then at each point
we remove a singular portion
from the Calabi--Yau whose boundary is topologically $S^2\times S^3$.
It is then possible to glue in $16$ copies of $B^3\times S^3$ to
obtain the smooth  quintic Calabi--Yau manifold. Note that the singular
 Calabi--Yau contains $\CP2$ as a smooth 4-manifold
passing through the $16$ singular points. When we remove the $16$ balls, we
remove
$16$ $B^4$'s from this $\CP2$. The boundary of each such $B^4$ is the $S^3$
we have glued in to desingularize the space, i.e.,
the vanishing cycles. The $\CP2$ with $16$ $B^4$'s removed
is thus a $4$-manifold-with-boundary on our Calabi--Yau manifold, and
its boundary is precisely the sum of all the vanishing cycles $\gamma^a$.
That is, \relation\ holds in the homology group.
Some relation such as \relation\ was to be expected from our count
of the codimension.

The singular quintics can alternatively be given a small resolution by
blowing up the $\CP2$ contained within them.  We blow up the locus
in $\CP4$ defined by $x_3=x_4=0$, which can be modeled inside
$\CP4\times\CP1$ as the set where $\{y_0\,x_4-y_1\,x_3=0\}$, $[y_0,y_1]$
being homogeneous coordinates on $\CP1$.  The original Calabi--Yau
conifold is blown up to a Calabi--Yau manifold defined by
\eqn\propert{\eqalign{y_0\,x_4-y_1\,x_3&=0\cr
y_0\, g(x) + y_1\, h(x)&=0 .}}

Topologically, this small resolution process glues in a copy
of $S^2\times B^4$
along each boundary $S^2\times S^3$.  There is a new $(1,1)$
class on the resolved space, which measures the area of the new
$S^2$'s which were added.  (All have the same area.)

\newsec{Quantum structure of conifolds}

In this section we will show that there is a smooth Wilsonian effective theory
at the conifold which includes light black holes, and that one can recover the
classical structure of conifolds (as described in section 2) by integrating out
these light black holes. The
argument follows \mbh, where further discussion can be found.

To make contact with the usual formulation of  $N=2,~~d=4$ supergravity and
special geometry \refs{\dlwp, \ascmp}, we introduce a symplectic basis of
three-cycles on the quintic
\eqn\trc{\langle A_I |B^J \rangle=-\langle B^J |A_I \rangle={\delta_I}^J,
{}~~~\langle A_I |A_J \rangle=\langle B^I |B^J \rangle=0,}
and corresponding periods
\eqn\zper{\eqalign{F_I&=\int_{A_I}\Omega,\cr Z^J&=\int_{B^J}\Omega.\cr}}
Only $30=15+15$ of the $204$ periods are relevant for our purposes, so we
henceforth restrict
\eqn\ifr{I,J=1,\cdots15.}
The $Z^K$ provide good local coordinates on $\cal M$, and
$\gamma^a$ may
be expanded
\eqn\sint{\gamma^a= {n^a}_IB^I.}
To be specific we choose a basis such that
\eqn\nex{\eqalign{{n^{16}}_I&=-1, \cr {n^a}_I&={\delta^a}_I,~~~a=1,...15.}}

For a compactification of type IIB string theory on $X$, the four dimensional
field theory has $101$ $N=2$ vector multiplets with associated
U(1)  field strengths which descend from the self-dual five form $F$ in ten
dimensions.
We are interested in the $15$ four-dimensional U(1) field strengths $G_I$
descended from $F$ as
\eqn\fss{F=(1+*)G_I\alpha^I,}
where $*$ is the ten-dimensional Hodge dual and $\alpha^I$ is a
harmonic three-form dual to $A_I$.

The ten-dimensional type IIB theory contains extremal black threebranes \bhole\
which can wrap around any one of the $16$ degenerating cycles $\gamma^a$
and appear as an extremal black hole in four dimensions. The mass $m^a$
of the $a$th such black
hole is determined from the Bogolmony bound \refs{\ferr} up to a constant
as
\eqn\mbb{m^a=|Z^a|=|{n^a}_IZ^I|,}
while the charge associated to the $I$th U(1) is
\eqn\qbh{{q^a}_I={n^a}_I.}
Evidently there are $16$ charged hypermultiplets  which can become
light as we approach a conifold point in the quintic moduli space. We shall
denote
these by $H^a$.  As in \refs{\swit,\mbh},
there is a smooth Wilsonian theory near the conifold
which
includes
the light fields $H^a$. Consider a generic region near a codimension one
conifold locus
$Z^K=0$ for which only one three-cycle $B^K$ vanishes. It is evident from \mbb\
that
only  one field
$H_K$ becomes light. Integrating out $H_K$ leads to effective couplings
between the vector multiplets which run (as a function of the moduli)
according to their
one-loop beta functions.
These running couplings are characterized by the holomorphic section \dlwp
\eqn\tua{\tau_{IJ}\sim {1 \over 2 \pi i}\delta_{IJ}\ln Z^J+ \hbox{\rm
(single-valued
function).}}
Using the special geometry relation
\eqn\tpr{\tau_{IJ}=\partial_IF_J,}
and integrating one finds the one-loop correction to
the
effective
periods near the hypersurface $Z^K=0$:
\eqn\flp{F_K \sim {1 \over 2 \pi i}Z^K\ln Z^K+ \hbox{\rm (single-valued
function)}, }
exactly as in \mbh. This agrees with the classical result  \asympt.

At  the
conifold locus where $\gamma^{16}$
degenerates, $\sum_K Z^K=0$ and the field $H^{16}$ becomes light. In the basis
\nex, this field
carries charge minus one with respect to each of the $15$ U(1)s. Consequently
there will be an identical contribution to the beta function of all
couplings, and $\tau$ will behave as
\eqn\tuba{\tau_{IJ}\sim {1 \over 2 \pi i}\ln \sum_{K=1}^{15}Z^K+ \hbox{\rm
(single-valued
function).}}
This implies a
monodromy
for each of the $15$ periods $F_L$ about this codimension one locus in
the moduli space:
\eqn\szr{F_L \sim {1 \over 2 \pi i}\sum_{J=1}^{15}Z^J\ln \sum_{K=1}^{15} Z^K+
\hbox{\rm (single-valued
function)}}
 Comparing with \asympt, we see that the monodromies computed by
integrating
out the light fields of the Wilsonian
effective field theory are in complete agreement with those of the classical
computation. The singularities of the classical moduli space
metric follow from \flp, \szr\ and relations of special geometry. Hence they
are also reproduced from the smooth Wilsonian theory.

It was crucial in our analysis that we regarded each of the $16$ states
obtained by wrapping threebranes around degenerating three-cycles as quanta of
independent quantum fields. This is motivated by consideration of the large
radius limit of the quintic conifold. In this limit the singular points are
widely
separated,
and the black holes correspond to well-localized threebranes wrapping around
degenerating three-cycles. Because they are well-separated it is natural to
quantize them as independent objects.
This critical assumption is an adaptation of the one made in \mbh, justified
in part by the overall consistency of the picture  presented here and
in the next section.
As pointed out in \mbh, it would be
desirable to find an independent assessment of the validity of this
method of counting states.

\newsec{Black hole condensation}

So far the story is similar in spirit, although more involved in detail, to the
case of the
simple conifold with a single vanishing cycle considered in \mbh. However
inspection of
the effective theory near the conifold reveals a dramatic new feature.
Each hypermultiplet $H^a$ contains two  charged
complex scalars which we
denote $h^{a\alpha}$ where $\alpha = 1,2$ is the global $SU(2)_R$ index of our
$N = 2$ representation.
 This gives a total of $32$ complex scalar fields.
Supersymmetry
implies a potential for these scalar fields of the general form
\refs{\dlwp,\west}
(taking the vevs of the scalar components of the vector multiplets to
vanish)
\eqn\actn{V\sim \sum_{I,J=1}^{15} M^{IJ}D_I^{\alpha \beta}D_{\alpha \beta J} }
where $M^{IJ}$
is a positive definite matrix \refs{\dlwp,\west} and
\eqn\dtr{ D_I^{\alpha \beta} = \sum_{a=1}^{16} q^a_I( h^{* a \alpha} h^{a
\beta} +
 h^{* a \beta} h^{a \alpha}).}
This potential has a flat direction along which
the three independent components of $D_I^{\alpha \beta}$ vanish,
\eqn\dfl{D_I^{\alpha \beta} = 0.}
This gives $45$ real constraints on the $32$ complex fields. In addition there
are $15$ gauge transformations which rotate the fields, leaving 4 real
vacuum parameters. Up to a
gauge transformation the general solution of \dfl\
in the basis \nex\ is
\eqn\flb{h^{a\alpha} = v^{\alpha}\qquad \hbox{for all $a$}}
for any complex two vector $v$.
Moving along the flat direction, the black holes condense and we
see that their moduli space
is parametrized by a single hypermultiplet.
The conifold point in the space of quintics (at which all $16$ cycles vanish)
corresponds to $v = 0$.
Moving away from this point along the flat direction corresponds to giving a
vev to
the charged hypermultiplets. This vev breaks all $15$ U(1)'s.
Thus we have discovered a second branch of the moduli space corresponding to a
charged black hole condensate. This branch has
$101-15=86$ massless vector multiplets, and $2+1=3$ massless hypermultiplets.

Compactification of  IIB string theory on a Calabi--Yau with
$(h_{21},h_{11})=(86, 2)$
leads to 86 vector multiplets and 3 hypermultiplets. A space with precisely
these Hodge
numbers arises if the singular conifold with $16$ degenerate cycles is resolved
with a $\CP1$, as discussed in section 3.
It is natural to identify the new branch of
the moduli space discovered in the analysis of the
Wilsonian effective field theory at the conifold with
compactification on the
$(86,2)$ Calabi--Yau. Further evidence for this identification can be obtained
by analyzing the behavior of the theory from the $(86,2)$ side and will
be presented in \bbgms.

While analysis of the general case will be deferred to later work,
one salient feature is worth mentioning. A general conifold has
$P$ vanishing cycles which obey $Q$ homology relations of the type
\relation. This implies $P$ degenerating black hole hypermultiplets
which carry charge with respect to $P-Q$ U(1)s. The generalization of the
potential \actn\ will then have $Q$ flat directions, since there are
$P-Q$ equations of the form \dfl\ for $P$ hypermultiplets. Generically all
$P-Q$ U(1)s will be broken along these flat directions.  Hence the new branch
of the moduli space will have $P-Q$ fewer vector multiplets and $Q$ more
hypermultiplets, corresponding to a Calabi-Yau space with Hodge numbers
$(h_{21}-P+Q, h_{11}+Q)$. This counting agrees precisely with one made
using the algebraic geometry methods of
\refs{\clemens,\schoen,\werner}
to compute the Hodge numbers of a Calabi-Yau space
obtained by degenerating and blowing up cycles.

\newsec{Mirror symmetry}

Two Calabi--Yau manifolds are said to constitute a mirror pair if
their corresponding conformally invariant nonlinear sigma models are isomorphic
via a mapping that flips the sign of an eigenvalue of
a $U(1)$ symmetry contained in the $N = 2$ superconformal algebra.
Such conformal theories can be used as the internal part of a string model,
and, as we have discussed, accurately describe string physics so long as we
are sufficiently far away from conifold points. We have learned in the above
discussion that even though the conformal field theory description of the
string model
breaks down at conifold points in the moduli space, the string description is
perfectly well behaved. Thus, as in \rgpmb, it seems worthwhile to
emphasize the notion of {\it string equivalence}:
two geometric
spaces are said to be string equivalent
 if they give rise to isomorphic models when
taken as the background spacetime for string theory. Mirror symmetry is
therefore
a special case of string equivalence in which the explicit isomorphism takes
the form noted above. More precisely, this isomorphism leads one to define,
in the context of type II string theory, two  Calabi--Yau spaces
as constituting a mirror pair if the type IIA string on the first is isomorphic
to the type IIB string on the second.
 Away from conifold points, where
conformal
field theory is valid, this is essentially
equivalent\foot{Some care must be taken concerning the integral
structures \udual.} to the standard formulation of mirror
symmetry.

Recall that in \gp\ a construction of pairs of mirror manifolds was presented
which relied crucially on the existence of special points in moduli space
at which the associated Calabi--Yau has enhanced discrete symmetries.
Nonetheless,
this construction was shown to establish the existence of mirror pairs
away from such special points through deformation arguments. Namely, if
$M$ and $W$ constitute a mirror pair at one point  in the moduli space, then
we can generate a family of mirrors by deforming, say, the complex structure of
$M$ and, correspondingly, the K\"ahler structure of $W$ (and vice versa).
The details of this notion were made precise in \agm, but the essential idea
is simple: since $M$ and $W$ give isomorphic physics, whatever operation is
performed on $M$ has a physically isomorphic description as an operation on
$W$,
thereby maintaining the mirror relationship.

This argument requires that the operation, say a deformation of the theory,
be smooth; otherwise we lose control and have no basis for drawing any
conclusions.
For this reason, such deformation arguments have only been used to establish
mirror symmetry for a continuously connected (in the sense of
conformal field theory) family of Calabi--Yau spaces
containing at least one point at which the explicit construction of
\gp\ could be applied. We now learn, from the results of the present work, that
we can continue such deformation arguments through conifold transitions since
string theory is perfectly well behaved along such a path. Thus, given
{\it one point}\/ in the web of connected Calabi--Yau spaces at which we
can establish the existence of a mirror partner (the construction of
\gp\ gives us many such points) we can use our deformation arguments to
establish
the existence of mirror symmetry at {\it all}\/ points in the web.
As almost
all
known Calabi--Yau manifolds are connected to the web
this establishes mirror
symmetry
for most, or possibly all,  Calabi--Yau spaces.\foot{The idea of using conifold
transitions to
yield a more general mirror construction was proposed some time ago
\morrison; the present
work is the physical realization of that mathematical conjecture.}

\bigskip

\centerline{\bf Acknowledgements}
We are grateful to
P. Aspinwall,
K. Becker, M. Becker, and J. Harvey
for useful discussions.
The work of B.R.G. was supported by
a National Young Investigator Award, the Alfred P. Sloan Foundation, and
a grant from the National Science Foundation.
The work of D.R.M. was supported in part by the United States Army
Research Office through the Mathematical Sciences Institute of
Cornell University, contract DAAL03-91-C-0027,
and in part by the National Science Foundation through
grants DMS-9401447 and PHY-9258582.
The work of A.S. was supported in part by DOE grant DOE-91ER40618.

\listrefs
\bye